\documentclass[runningheads]{llncs}
\usepackage[T1]{fontenc}
\usepackage{graphicx}
\usepackage{multirow}
\usepackage{multicol}
\usepackage{booktabs}

\usepackage{hyperref} 
\usepackage{cleveref} 

\hypersetup{colorlinks = false, 
	    linkcolor = black, 
	    urlcolor = blue,
            citecolor = blue} 

\usepackage{marvosym}

\begin{document}
\title{CAVM: Conditional Autoregressive Vision Model for Contrast-Enhanced Brain Tumor MRI Synthesis}
\titlerunning{CAVM for Contrast-Enhanced Brain Tumor MRI Synthesis}
\author{Lujun Gui\inst{1} \and
Chuyang Ye\inst{2} \and
Tianyi Yan\inst{1}\textsuperscript{(\Letter)}}
\authorrunning{L. Gui et al.}
\institute{School of Medical Technology, Beijing Institute of Technology, Beijing, China \\
\email{yantianyi@bit.edu.cn}
\and
School of Integrated Circuits and Electronics, Beijing Institute of Technology, Beijing, China}
\renewcommand{\thefootnote}{}
\footnotetext{L.Gui and C.Ye--Equal contribution.}
\maketitle           
\begin{abstract}
Contrast-enhanced \textit{magnetic resonance imaging} (MRI) is pivotal in the pipeline of brain tumor segmentation and analysis.
Gadolinium-based contrast agents, as the most commonly used contrast agents, are expensive and may have potential side effects, and it is desired to obtain contrast-enhanced brain tumor MRI scans without the actual use of contrast agents.
Generic deep learning methods have been applied to synthesize virtual contrast-enhanced MRI scans from non-contrast images.
However, as this synthesis problem is inherently ill-posed, these methods fall short in producing high-quality results.
In this work, we propose \textit{Conditional Autoregressive Vision Model} (CAVM) for improving the synthesis of contrast-enhanced brain tumor MRI.
As the enhancement of image intensity grows with a higher dose of contrast agents, we assume that it is less challenging to synthesize a virtual image with a lower dose, where the difference between the contrast-enhanced and non-contrast images is smaller.
Thus, CAVM gradually increases the contrast agent dosage and produces higher-dose images based on previous lower-dose ones until the final desired dose is achieved.
Inspired by the resemblance between the gradual dose increase and the Chain-of-Thought approach in natural language processing, CAVM uses an autoregressive strategy with a decomposition tokenizer and a decoder.
Specifically, the tokenizer is applied to obtain a more compact image representation for computational efficiency, and it decomposes the image into dose-variant and dose-invariant tokens.
Then, a masked self-attention mechanism is developed for autoregression that gradually increases the dose of the virtual image based on the dose-variant tokens.
Finally, the updated dose-variant tokens corresponding to the desired dose are decoded together with dose-invariant tokens to produce the final contrast-enhanced MRI.
CAVM was validated on the publicly available BraSyn-2023 dataset with brain tumor MRI scans, where it outperforms several state-of-the-art methods for medical image synthesis. 
The code is available at \href{https://github.com/Luc4Gui/CAVM}{https://github.com/Luc4Gui/CAVM}.

\end{abstract}
\section{Introduction}
Gadolinium-based contrast agents significantly improve the diagnostic accuracy based on brain \textit{magnetic resonance imaging} (MRI) by enhancing the contrast of vascular structures and pathological conditions~\cite{zhou2013gadolinium}.
The contrast-enhanced brain MRI scan is crucial in detecting and characterizing brain tumors~\cite{baid2021rsna}.
However, the use of gadolinium-based contrast agents requires careful consideration of potential risks, such as nephrogenic systemic fibrosis in patients with renal impairment and gadolinium deposition in the brain~\cite{schieda2018gadolinium}.
To avoid the use of contrast agents, cross-modality MRI synthesis models are developed to estimate contrast-enhanced brain tumor MRI scans from available non-contrast modalities~\cite{dayarathna2023deep}.

Most current cross-modality MRI synthesis models~\cite{jiang2023cola,liu2023one} are based on deep learning, and they are trained with paired input source and output target modality images.
For example, ResViT~\cite{dalmaz2022resvit} is a \textit{generative adversarial network}~(GAN) approach for cross-modality MRI synthesis, which harnesses the contextual sensitivity of vision Transformers~\cite{dosovitskiy2020image}, combined with the precision of convolutional operators and the authenticity encouraged by adversarial learning; 
SynDiff~\cite{ozbey2023unsupervised} is based on the diffusion model~\cite{ho2020denoising}, where a conditional diffusion process is used and large diffusion steps are replaced by adversarial projections for computational efficiency;
the task-specific sequence-to-sequence network TSF-Seq2Seq~\cite{han2023explainable} transforms cross-modality MRI synthesis from an image-to-image problem into a sequence-to-sequence problem by fusing multi-modal information;
MT-Net~\cite{li2023multi} is a cross-modality MRI synthesis framework with edge-aware self-supervised pre-training, which can alleviate the challenge of limited paired data.
However, the performance of these methods is still unsatisfactory for contrast-enhanced brain tumor MRI synthesis from non-contrast MRI, as the problem is inherently ill-posed due to the lack of contrast in non-contrast MRI.

Unlike general cross-modality MRI synthesis, the problem of contrast-enhanced MRI synthesis features a unique continuous variable, the contrast agent dose, where the difference between contrast-enhanced MRI and non-contrast MRI is greater with an increasing dose~\cite{zhou2013gadolinium}.
Yet this property has not been explored by previous works on the contrast-enhanced MRI synthesis problem.
Based on this relationship between image contrast and the contrast agent dose, we assume that it is less challenging to synthesize a virtual contrast-enhanced image with a lower dose and seek to accordingly further explore the challenging problem of contrast-enhanced brain tumor MRI synthesis from non-contrast MRI.

Specifically, we propose the \textit{Conditional Autoregressive Vision Model} (CAVM), which synthesizes contrast-enhanced MRI by gradually increasing the dose until the final desired dose is achieved. 
Motivated by the resemblance between the gradual dose increase and the Chain-of-Thought approach~\cite{wei2022chain} in natural language processing, CAVM uses an autoregressive strategy with a decomposition tokenizer and a decoder.
In CAVM, the tokenizer is first applied to the input images to obtain a more compact image representation for computational efficiency, and it decomposes the image into dose-variant and dose-invariant tokens.
Then, a masked self-attention mechanism is developed based on LLaMA-style~\cite{touvron2023llama} Transformer blocks for autoregression, which gradually produces images of increased doses with the dose-variant tokens of non-contrast images and intermediate lower-dose images that are already produced. 
Finally, the updated dose-variant tokens associated with the desired dose are decoded together with dose-invariant tokens to produce the final contrast-enhanced MRI.
CAVM was validated on the publicly available BraSyn-2023 dataset~\cite{li2023brain} and compared with state-of-the-art medical image synthesis methods.
The results show that our method improves the synthesis quality by noticeable margins.

\section{Methodology}

\subsection{Problem Formulation and Method Overview}
Given non-contrast brain tumor MRI scans of different modalities, including T1-weighted~(T1w), T2-weighted (T2w), and \textit{fluid-attenuated inversion recovery} (FLAIR) images, our work aims to synthesize a corresponding T1 gadolinium-based contrast-enhanced (T1Gd) image, so that contrast-enhanced brain tumor MRI can be obtained without actually injecting contrast agents.
We also assume that the brain tumor mask is available for assistance, where a model can be trained to reliably perform brain tumor segmentation on the T1w, T2w, and FLAIR images.
For convenience, we denote the non-contrast images by ${x}_{\mathrm{NC}}$, the tumor mask by ${x}_{\mathrm{TM}}$, and the standard-dose T1Gd image by ${y}_{\mathrm{SD}}$.
By concatenating ${x}_{\mathrm{NC}}$ and ${x}_{\mathrm{TM}}$, we obtain ${x}$, which is the overall input to the synthesis model.

\begin{figure}[!t]
\includegraphics[width=\textwidth]{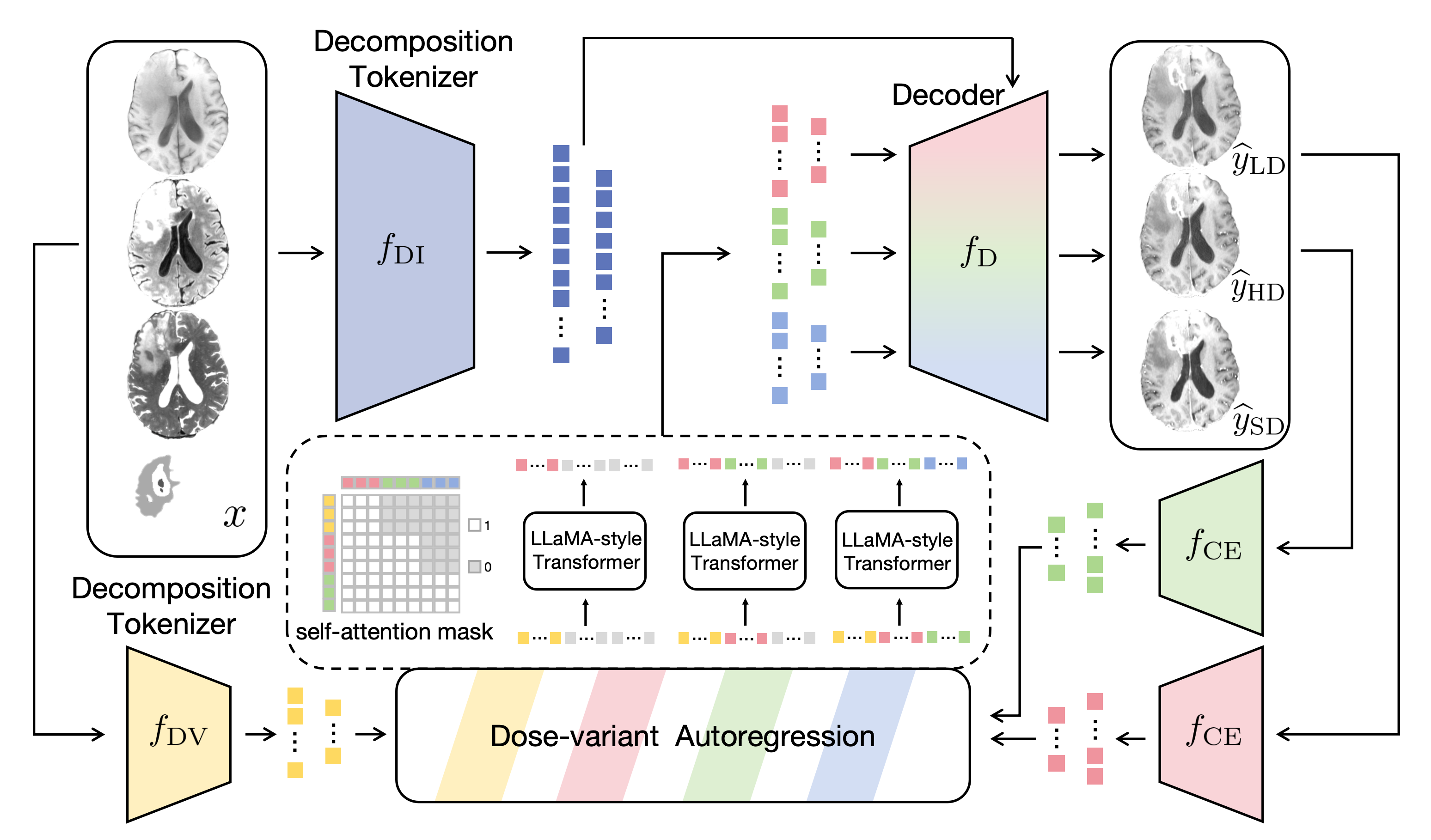}
\caption{The overall architecture of CAVM. Decomposition Tokenizer comprises two encoders located on the left. The Dose-variant Autoregression is implemented by LLaMA-style Transformer and two encoders situated at the bottom right. The decoder, positioned at the top right of the diagram, decodes all output images from image tokens during the autoregressive process.
} \label{fig1}
\end{figure}

To reduce the difficulty in synthesizing contrast-enhanced brain tumor MRI from non-contrast MRI, we transform the synthesis of contrast-enhanced information into three sequential steps, where contrast-enhanced information is gradually recovered, first for a lower dose, then for a higher dose, and finally for the standard dose.
We propose an autoregressive model CAVM for the gradual dose increase, the design of which is illustrated in Fig.~\ref{fig1}.
The input images are tokenized first and decomposed into dose-variant and dose-invariant tokens.
The dose-variant tokens are fed into an autoregression module that gradually increases the dose.
Then, the updated dose-variant tokens are decoded with the dose-invariant tokens to produce the target contrast-enhance brain tumor MRI.
The detailed design of the model architecture is described below.

\subsection{Model Architecture}

\subsubsection{Decomposition Tokenizer}
Since contrast agents do not alter the intensities of all image regions, we decompose the input images into a dose-variant component and a dose-invariant component, so that the synthesis problem can focus on the relevant part and its difficulty can be reduced.
Moreover, as the dimension of the original image is high, for computational efficiency, the dose-variant and dose-invariant components are represented by lower-dimensional tokens. 
The dose-variant and dose-invariant tokens are obtained with a decomposition tokenizer, which is built upon the encoder of Swin UNETR~\cite{he2023swinunetr}, a backbone proven powerful in both medical image segmentation and generation, to ensure the quality of the tokens. 
The decomposition tokenizer comprises two distinct units, ${f}_{\mathrm{DV}}$ for dose-variant tokenization and ${f}_{\mathrm{DI}}$ for dose-invariant tokenization.
They comprise two and four stages of the Swin UNETR encoder, respectively, and extract the outputs of the last two stages as tokens. 

\subsubsection{Dose-variant Autoregression}

The autoregression module then applies gradual dose increase by autoregressing the dose-variant tokens, where tokens associated with higher doses are obtained sequentially.
Multiple steps are used in the autoregression, where at each step, a higher-dose token is obtained based on the non-contrast dose-variant tokens and all lower-dose tokens that have been obtained, until the desired dose is reached.
Here, we choose to use three autoregression steps, which balance the computation complexity and reduction in regression difficulty.
This leads to the sequential synthesis of a lower-dose image ${y}_{\mathrm{LD}}$, a higher-dose image ${y}_{\mathrm{HD}}$, and the standard-dose image ${y}_{\mathrm{SD}}$.

We construct the autoregression module based on LLaMA-style Transformer blocks with RMSNorm normalization and rotary position embedding~\cite{touvron2023llama}, as they have achieved excellent performance among Transformer-based architectures in both computer vision and natural language processing~\cite{li2024llava,touvron2023llama2}.
Mathematically, suppose the dose-variant tokens of ${x}$ given by $f_{\mathrm{DV}}$ is $t_{x}$, and the dose-variant tokens of ${y}_{\mathrm{LD}}$, ${y}_{\mathrm{HD}}$, and ${y}_{\mathrm{SD}}$ are $t_{\mathrm{LD}}$, $t_{\mathrm{HD}}$, and $t_{\mathrm{SD}}$, respectively; then the autoregression is formulated as
\begin{eqnarray}
    t_{\mathrm{LD}},t_{0},t_{0} &=&\mathrm{MMHSA}\left( \mathrm{RoPE}\left( W\times \left( t_{x},t_{0},t_{0}\right) \right) \right) + \left( t_{x},t_{0},t_{0}\right) \\
    t_{\mathrm{LD}},t_{\mathrm{HD}},t_{0} &=&\mathrm{MMHSA}\left( \mathrm{RoPE}\left( W\times \left( t_{x},\tilde{t}_{\mathrm{LD}},t_{0}\right) \right) \right) + \left( t_{x},\tilde{t}_{\mathrm{LD}},t_{0}\right) \\
    t_{\mathrm{LD}},t_{\mathrm{HD}},t_{\mathrm{SD}} &=&\mathrm{MMHSA}\left( \mathrm{RoPE}\left( W\times \left( t_{x},\tilde{t}_{\mathrm{LD}},\tilde{t}_{\mathrm{HD}}\right) \right) \right) + \left( t_{x},\tilde{t}_{\mathrm{LD}},\tilde{t}_{\mathrm{HD}}\right)
\end{eqnarray}
Here, $t_0$ represents placeholder tokens, $W$ is a linear weighting layer, $\mathrm{RoPE}(\cdot)$ denotes rotary position embedding~\cite{su2024roformer}, and $\mathrm{MMHSA}(\cdot)$ denotes the masked multi-head self-attention~\cite{vaswani2017attention}; moreover, instead of using the autoregressed $t_{\mathrm{LD}}$ and $t_{\mathrm{HD}}$ directly for subsequently autogression steps, we use their updated versions, $\tilde{t}_{\mathrm{LD}}$ and $\tilde{t}_{\mathrm{HD}}$ respectively.
$\tilde{t}_{\mathrm{LD}}$ and $\tilde{t}_{\mathrm{HD}}$ are obtained by feeding $t_{\mathrm{LD}}$ and $t_{\mathrm{HD}}$ into a decoder for image reconstruction (to be introduced next) and then using a shared tokenizer ${f}_{\mathrm{CE}}$ to encode the reconstructed low-dose and high-dose contrast-enhanced images, respectively.
This update allows dose-invariant tokens to correct the potential errors in autoregressed dose-variant tokens.
${f}_{\mathrm{CE}}$ comprises four stages of the Swin UNETR encoder and gives the outputs of the two last layers. 

Note that in CAVM, we revise the original design of the self-attention mask in MMHSA, as the default lower triangular self-attention mask only uses the tokens in the previous state.
To fully exploit all tokens of previous states, the self-attention mask is designed with a staircase shape (also shown in Fig.~\ref{fig1}).
Specifically, the $i$-th token $z_{i}$ in $(t_{\mathrm{LD}},t_{\mathrm{HD}},t_{\mathrm{SD}})$ is computed as
\begin{equation}
    z_{i}=\sum ^{K}_{k=1}a_{ik}v_{k},
\end{equation}
where $K$ is the total number of tokens, $v_{k}$ is the value embedding, and $a_{ik}$ is the attention weight associated with $v_{k}$.
Suppose $n$ is the number of tokens per image; then the masking scheme is achieved by setting $a_{ik} = 0$ for $\lfloor k/n \rfloor > \lfloor i/n \rfloor$ and $\sum ^{K}_{k=1}a_{ik} = 1$.
Another important benefit of the revised masking is the inference speedup.
With this staircase masking, the whole image can be processed with a single forward pass instead of one forward pass for each single token, which is computationally impractical for our task.

\subsubsection{Decoder for Reconstruction}

To obtain the contrast-enhanced images from the tokens with gradually increased doses, a single decoder ${f}_{\mathrm{D}}$ combines the dose-invariant tokens $t_{x}$ of the input images with $t_{\mathrm{LD}}$, $t_{\mathrm{HD}}$, and $t_{\mathrm{SD}}$ and produces the estimated low-dose, high-dose, and standard-dose images $\hat{y}_{\mathrm{LD}}$, $\hat{y}_{\mathrm{HD}}$, and $\hat{y}_{\mathrm{SD}}$, respectively.
${f}_{\mathrm{D}}$ uses four stages of the decoder of Swin UNETR~\cite{he2023swinunetr} for image reconstruction.

\subsection{Model Training}
We first pretrain the tokenizer $f_{\mathrm{DI}}$, $f_{\mathrm{DV}}$, and $f_{\mathrm{CE}}$ with two joint tasks.
The first task is an autoencoding task for contrast-enhanced images ${y}_{\mathrm{LD}}$, ${y}_{\mathrm{HD}}$, and ${y}_{\mathrm{SD}}$, where $f_{\mathrm{DI}}$ and $f_{\mathrm{CE}}$ are involved.
Note that although the training data only provides non-contrast images and standard-dose contrast-enhanced brain tumor MRI, it is possible to generate synthetic lower-dose (33\% dose) and higher-dose~(66\% dose) images from them for autoencoding with the approach in~\cite{pinetz2023faithful}.
The second task is an image-to-image prediction task from ${x}$ to ${y}_{\mathrm{SD}}$, where $f_{\mathrm{DI}}$ and $f_{\mathrm{DV}}$ are involved. 
For each task, the L1 loss and adversarial loss are used for training, where we use the discriminator from~\cite{dalmaz2022resvit} and extend it to 3D.
The final training loss is the direct sum of the two losses.
Then, we train the dose-variant autoregression module with
the $l_{2}$ regression loss from AIM~\cite{el2024scalable} and MAE~\cite{he2022masked} based on the predicted lower-dose, higher-dose, standard-dose images. 

The hyperparameter settings of the LLaMA-style Transformer blocks are as follows. We set the embedding dimensions to 384 and 768, multi-attention heads to 4 and 8, LLaMA-style Transformer layers to 4 and 8, and maximum sequence lengths to 432 and 864. 
Model training was performed with Adam ($\beta_1=0.9$ and $\beta_2=0.99$)~\cite{kingma2014adam}, where we set the batch size to 1 and peak learning rate to $10^{-4}$.

\section{Experiments and Results}
\subsection{Experimental Setups}
We evaluated CAVM on the public BraSyn-2023 dataset~\cite{li2023brain}, which comprised aligned multi-contrast brain tumor MRI scans of 1,470 subjects, as well as the tumor annotations.
The image modalities included the T1w, T2w, FLAIR, and T1Gd images.
We randomly selected 1,043, 219, and 208 subjects as the training, validation, and test set, respectively.
All images were center-cropped to $192\times192\times128$ and normalized by dividing the image intensities by the value at the 95th percentile of the intensity distribution.
To provide the tumor mask input to CAVM, a nnU-Net model~\cite{isensee2021nnu} was trained with the T1w, T2w, and FLAIR images and annotations in the training set.

We computed the \textit{Structural Similarity Index Measure} (SSIM) and \textit{Peak Signal-to-Noise Ratio} (PSNR) to assess image quality.
These two metrics were summarized for the tumor region and other healthy brain tissue separately.
To assess the utility of CAVM for downstream analysis, brain tumor segmentation was performed with another nnU-Net model trained with the T1w, T2w, FLAIR, and T1Gd images and annotations in the training set.
This model was applied to the real T1w, T2w, and FLAIR images and synthesized T1Gd images of the test set, and the Dice coefficient was computed for the segmentation result.

\subsection{Performance Evaluation}

\begin{figure}[!t]
\includegraphics[width=\textwidth]{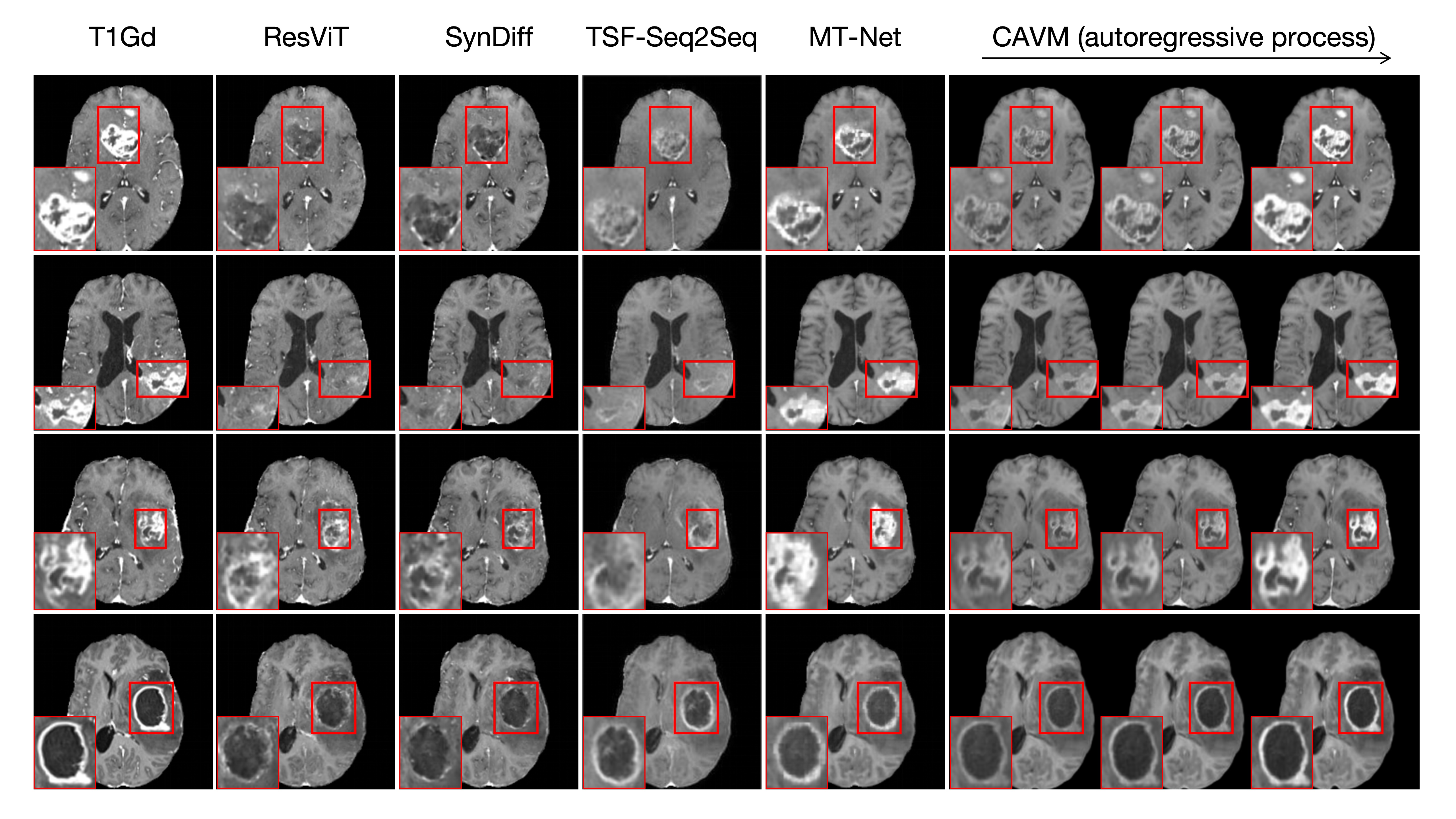}
\caption{Four examples of real T1Gd images and synthesized results. For CAVM, from left to right the image order is lower-dose, higher-dose, and standard-dose. Note the highlighted tumor region for comparison.} \label{fig2}
\end{figure}

We compared CAVM with four state-of-the-art models for multi-contrast brain MRI synthesis, including a GAN-based method ResViT~\cite{dalmaz2022resvit}, a diffusion model SynDiff~\cite{ozbey2023unsupervised}, a sequence-to-sequence model TSF-Seq2Seq~\cite{han2023explainable}, and the autoencoder pretrained MT-Net~\cite{li2023multi}.
For a fair comparison, for models (ResViT, SynDiff, and MT-Net) that can incorporate brain tumor masks into the input, their inputs were identical to those of CAVM; otherwise, the input only comprised the non-contrast images.

Fig.~\ref{fig2} presents examples of synthesized T1Gd images produced by different models.
ResViT and SynDiff failed to reproduce the enhanced intensity in the tumor.
TSF-Seq2Seq and MT-Net produced enhanced intensities, but the results are not consistent with the real images.
By contrast, CAVM generated accurate T1Gd images with faithful contrast enhancement in the tumor.
Also, the intermediate lower-dose and higher-dose results confirm that CAVM gradually increased the dose for the synthesized image.

Table~\ref{tab1} quantitatively compared CAVM with the competing methods.
The results show that CAVM outperforms the other models in both tumor regions and healthy tissue.
In the tumor region, CAVM surpasses the other models with a relative improvement of at least 9.61\% in PSNR, 5.40\% in SSIM, and 5.39\% in the Dice coefficient.
In healthy tissue, CAVM also has better performance than the competing methods, although the margin is smaller as synthesis of healthy tissue is less challenging.
Note that the upper-bound mean Dice coefficient of tumor segmentation achieved on real images is $90.19$ (\%), and the Dice coefficient achieved with the synthesized image of CAVM is 88.74, which is close to the upper bound. 

\begin{table}[!t]
\centering
\caption{The SSIM (\%), PSNR (dB), and Dice coefficient (\%) results (mean$\pm$std) of each method on the test set. Boldface marks the top result.}\label{tab1}
\begin{tabular}{cccccccccccccccc}
\toprule
\hline
\multirow{2}{*}{Method}     & & & & \multicolumn{5}{c}{Brain Tumor}                  & & & & \multicolumn{3}{c}{Healthy Tissue} & \cr
\cline{5-9} \cline{13-15}
                            & & & & SSIM           & & PSNR           & & Dice           & & & & SSIM           & & PSNR &           \cr
\hline
ResViT                & & & & 88.01$\pm$0.53 & & 18.16$\pm$2.97 & & 74.25$\pm$9.83 & & & & 93.97$\pm$0.20 & & 27.24$\pm$2.00 & \\
SynDiff               & & & & 88.12$\pm$0.55 & & 18.21$\pm$2.94 & & 74.14$\pm$9.65 & & & & 93.62$\pm$0.19 & & 26.91$\pm$1.96 & \\
TSF-Seq2Seq           & & & & 86.03$\pm$0.72 & & 19.71$\pm$3.28 & & 82.25$\pm$9.12 & & & & 93.10$\pm$0.37 & & 27.35$\pm$2.15 & \\
MT-Net                & & & & 89.64$\pm$0.66 & & 23.82$\pm$3.42 & & 84.20$\pm$9.89 & & & & 95.04$\pm$0.17 & & 32.38$\pm$1.80 & \\
CAVM                  & & & & \textbf{94.48$\pm$0.35} & & \textbf{26.11$\pm$2.87} & & \textbf{88.74$\pm$2.86} & & & & \textbf{95.50$\pm$0.19} & & \textbf{32.91$\pm$1.94} & \\
\hline
\bottomrule
\end{tabular}
\end{table}

\begin{table}[!t]
\centering
\caption{The SSIM (\%) and PSNR (dB) results (mean$\pm$std) of the ablation study. Boldface marks the top result.}\label{tab2}
\begin{tabular}{cccccccc}
\toprule
\hline
\multirow{2}{*}{Method}           & & & & \multicolumn{3}{c}{Brain Tumor}                  & \cr
\cline{5-7}
                                    & & & & SSIM           & & PSNR           & \cr
\hline
One Step                     & & & & 89.98$\pm$0.77 & & 24.22$\pm$3.69 & \\
Two Steps                     & & & & 92.53$\pm$0.65 & & 24.92$\pm$3.56 & \\
w/o Dose-variant Autoregression   & & & & 89.70$\pm$0.70 & & 23.97$\pm$3.91 & \\
CAVM                     & & & & \textbf{94.48$\pm$0.35} & & \textbf{26.11$\pm$2.87} & \\
\hline
\bottomrule
\end{tabular}
\end{table}

\subsection{Ablation Study}
We performed an ablation study to confirm the contribution of each major component of CAVM by individually removing them.
We first investigated the benefit of each dose increase step in the autoregression, where one and two steps were considered.
With one step, CAVM synthesized the T1Gd image without considering lower-dose and higher-dose images, whereas with two steps, the synthesis only considered the lower-dose image.
Also, we have completely removed the dose-variant autoregression module, where the T1Gd image was directly obtained by decoding the outputs of the tokenizer.
Note that the removal of the decomposition that directly increases the dose based on all image tokens is computationally infeasible, and thus it was not considered.
The SSIM and PSNR results in the tumor regions are summarized for the ablation study in Table~\ref{tab2}.
Fewer steps led to worse image qualities, and without the autoregression the synthesis performance also decreased. These results show that the dose increase steps and the complete autoregression module are all beneficial to the synthesis.

\section{Conclusion}
In this work, we present CAVM, a conditional autoregressive vision model for contrast-enhanced brain tumor MRI synthesis from non-contrast images.
CAVM gradually increases the dosage of the contrast agent and produces higher-dose images from previous lower-dose ones until the final desired dose is achieved. 
The dose increase is achieved with autoregression based on LLaMA-style Transformer blocks and dose-variant and dose-invariant token decomposition.
Experimental results on public data show that CAVM compares favorably with state-of-the-art synthesis models, particularly in tumor regions, and it also improves downstream brain tumor segmentation.
It is worth noting that CAVM does not sample the output hidden state, which is an issue that we will further focus on in future work.

\begin{credits}
\subsubsection{\ackname} C. Ye is supported by the Beijing Municipal Natural Science Foundation (7242273), and the Fundamental Research Funds for the Central Universities (2024CX06040). 
T. Yan is co-founded by Key-Area Research and Development Program of Guangdong Province (2023B0303030002), the STI 2030-Major Projects (grant number 2022ZD0208500), and the National Natural Science Foundation of China (grant numbers U20A20191, 62336002).

\subsubsection{\discintname}
The authors have no competing interests to declare that are relevant to the content of this article.
\end{credits}
%
%
%
\bibliographystyle{splncs04}
\bibliography{mybibliography}
\end{document}